\documentclass[aps,floatfix,pra,superscriptaddress,reprint,showpacs,10pt,preprintnumbers,longbibliography]{revtex4-2}
\usepackage[utf8]{inputenc}
\usepackage[pdftex]{graphicx}
\usepackage{float}
\usepackage{amssymb}
\usepackage{amsmath}
\usepackage{dsfont}
\usepackage{array}
\usepackage{bm}
\usepackage{subfigure}
\usepackage{mathrsfs}
\usepackage{pifont}
\usepackage{multirow}
\usepackage{upgreek}
\usepackage{xcolor}
\usepackage[pdftex,
            pdftitle={Exploring the CP-Violating Dashen Phase in the Schwinger Model with Tensor Networks},
            pdfauthor={Lena Funcke, Karl Jansen, Stefan Kuehn},
            bookmarks,
            colorlinks,            
            menucolor=black,            
            plainpages=false,
            linkcolor=myblue,
            citecolor=mymagenta,
            menucolor=black,
            urlcolor=myblue,
            pdfpagelabels,
            hypertexnames=false]{hyperref}
\usepackage{verbatim}
\usepackage{slashed}
\usepackage{cleveref}
\usepackage[braket,qm]{qcircuit}
\usepackage[T1]{fontenc}
\DeclareTextSymbolDefault{\dh}{T1}
\usepackage{hyperref}

\definecolor{mymagenta}{RGB}{200, 0, 100}
\definecolor{myblue}{RGB}{45, 48, 146}

\newcommand{\tr}{\ensuremath{\text{tr}}}

\graphicspath{{figures/}}

\begin{document}
\title{Exploring the CP-violating Dashen phase in the Schwinger model with tensor networks}
\author{Lena Funcke}
\affiliation{Transdisciplinary Research Area ``Building Blocks of Matter and Fundamental Interactions'' (TRA Matter) and Helmholtz Institute for Radiation and Nuclear Physics (HISKP), University of Bonn, Nußallee 14-16, 53115 Bonn, Germany}
\affiliation{Center for Theoretical Physics, Co-Design Center for Quantum Advantage, and NSF AI Institute for Artificial Intelligence and Fundamental Interactions, Massachusetts Institute of Technology, 77 Massachusetts Avenue, Cambridge, MA 02139, Cambridge, MA, USA}
\author{Karl Jansen}
\affiliation{DESY Zeuthen, Platanenallee 6, 15738 Zeuthen, Germany}
\author{Stefan K{\"u}hn}
\affiliation{DESY Zeuthen, Platanenallee 6, 15738 Zeuthen, Germany}
\affiliation{Computation-Based Science and  Technology Research Center, The Cyprus  Institute, 20 Kavafi Street, 2121 Nicosia, Cyprus}

\begin{abstract}
We numerically study the phase structure of the two-flavor Schwinger model with matrix product states, focusing on the (1+1)-dimensional analog of the CP-violating Dashen phase in QCD. We simulate the two-flavor Schwinger model around the point where the positive mass of one fermion flavor corresponds to the negative mass of the other fermion flavor, which is a sign-problem afflicted regime for conventional Monte Carlo techniques. Our results indicate that the model undergoes a CP-violating Dashen phase transition at this point, which manifests itself in abrupt changes of the average electric field and the analog of the pion condensate in the model. Studying the scaling of the bipartite entanglement entropy as a function of the volume, we find clear indications that this transition is not of first order.
\end{abstract}

\maketitle

\section{Introduction\label{sec:introduction}}
 
Monte Carlo (MC) methods have shown unparalleled success for exploring static properties of lattice gauge theories, such as mass spectra~\cite{Durr2008,Alexandrou2014,Bali2022} and phase diagrams~\cite{Guenther2021}. However, conventional MC methods cease to work in certain parameter regimes, in which the Euclidean lattice action of the theory becomes negative or complex. This prevents an efficient MC sampling, and is generally referred to as sign problem. 

A prominent example is quantum chromodynamics (QCD) in the presence of a large topological $\theta$-term. Here, a particularly interesting point in parameter space is $\theta=\pi$, where a CP-violating phase transition has been predicted in early works by Dashen using current algebra arguments~\cite{Dashen1971} and later using anomaly matching techniques~(see Ref.~\cite{Gaiotto2017} and references therein). Going beyond QCD, the Dashen phase transition also plays a crucial role in several models beyond the Standard Model of particle physics (see, e.g., Refs.~\cite{DiVecchia:2017xpu,Perez:2020dbw}).

\begin{figure}[tb!]
  \centering
  \includegraphics[width=0.35\textwidth]{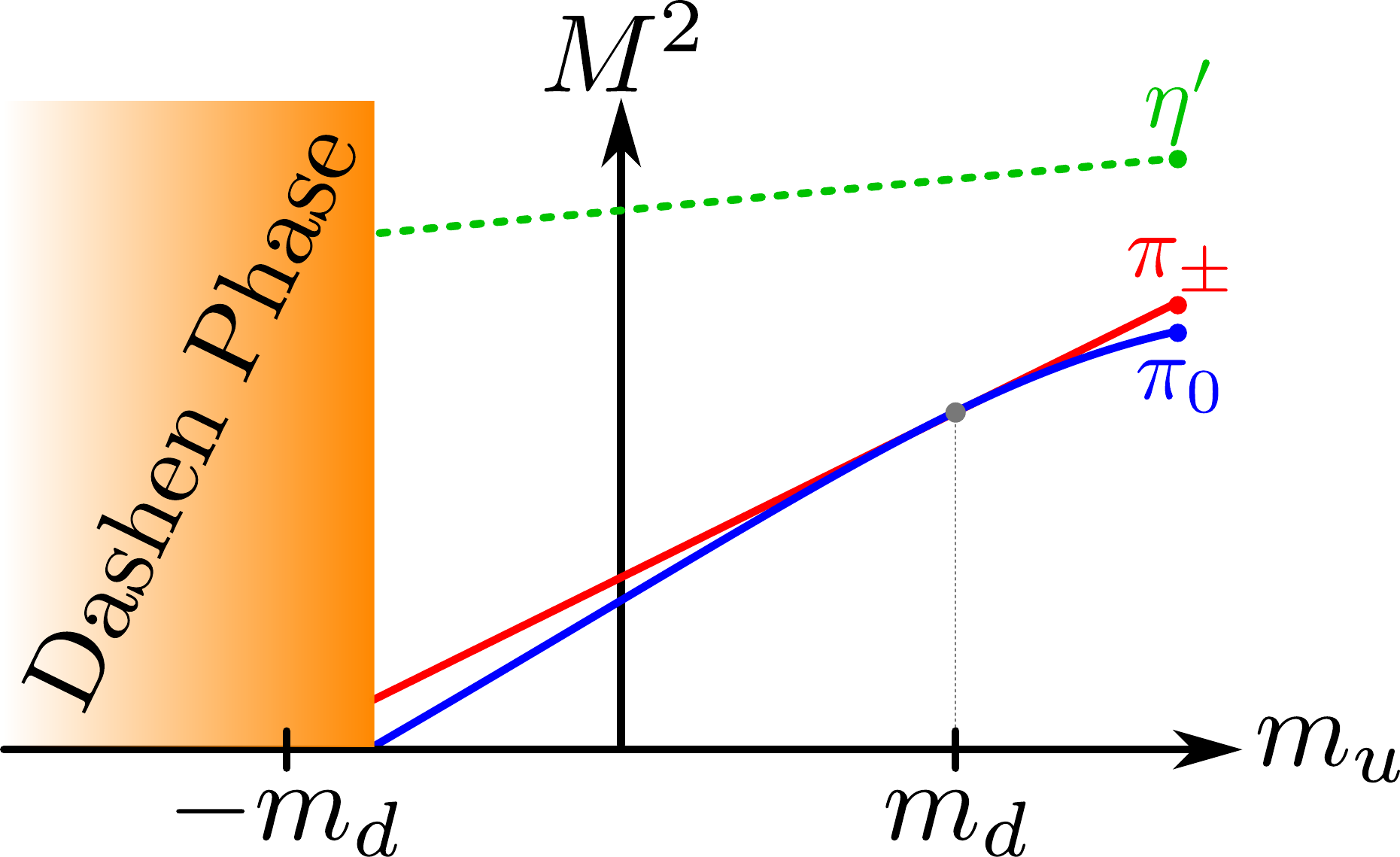}
  \caption{Illustration of the low-lying hadronic spectrum of two-flavor QCD. Fixing the down-quark mass $m_d$ to a positive value and decreasing the up-quark mass $m_u$ to negative values, the neutral pion mass $M_{\pi_0}^2\propto m_u+m_d$ eventually becomes complex, which indicates the onset of the Dashen phase transition. The transition is expected to happen for values of $m_u$ slightly larger than $-m_d$ because of corrections $\propto (m_u-m_d)^2$ to the neutral pion mass.}
  \label{fig:PhaseDiagram}
\end{figure}

The origin of the Dashen phase transition can be easily understood by noting that $\theta=\pi$ corresponds to a negative quark mass, due to the chiral anomaly. Following Refs.~\cite{Creutz:1995wf,Creutz2013,Creutz2014,Creutz2019}, let us consider QCD with only the lightest two fermion flavors, i.e., the up- and the down-quark. The low-lying hadronic spectrum is given by the pseudoscalar mesons, i.e., the pions and the two-flavor analog of the $\eta'$ meson. Neglecting electromagnetic effects, chiral perturbation theory predicts that the pion mass squared, $M_\pi^2$, is proportional to $m_u + m_d$, where $m_u$ ($m_d$) is the bare mass of the up (down) quark. Due to corrections $\propto (m_u-m_d)^2$~\cite{Gasser1983,Gasser1984}, the neutral pion $\pi_0$ has a slightly smaller mass than the charged pions $\pi_\pm$ (see Fig.~\ref{fig:PhaseDiagram}). Choosing a fixed positive value of the down-quark mass, $m_d>0$, the neutral pion mass is positive as long as $m_u\gtrsim -m_d$, and hence the theory is gapped in this regime. When decreasing $m_u$ further, the neutral pion mass vanishes and eventually becomes complex, thus indicating the onset of the Dashen phase transition, as illustrated in Fig.~\ref{fig:PhaseDiagram}. At the transition, the pion condenses and acquires a nonzero vacuum expectation value. This transition spontaneously breaks the CP symmetry, because the pion is a CP-odd particle~\cite{Creutz:1995wf,Creutz2013,Creutz2014,Creutz2019}.

The order of the Dashen phase transition depends on an unknown sign in the effective action for the two-flavor case. Reference~\cite{Creutz:1995wf} argued that a series of first-order transition lines should exist along the $m_d$-axis, which end in second-order transitions. However, this conjecture has not been verified numerically. 

In this paper, we numerically study the Dashen phase transition in the two-flavor Schwinger model~\cite{Schwinger1962,Coleman1975,Coleman1976}, which is a model that shares many properties with QCD and has therefore been adopted as a benchmark model for testing new numerical methods aimed at QCD applications.  Our study is an extension of our previous preliminary investigation of the Dashen phase transition in Ref.~\cite{Funcke:2021glr}. In our study, we adopt a Hamiltonian lattice formulation with Kogut-Susskind staggered fermions and simulate the model with matrix product states (MPS), a particular kind of tensor network (TN). With MPS, the spectrum of the single-flavor Schwinger model has already been successfully computed~\cite{Banuls2013,Banuls2013a,Buyens2013,Kuehn2014,Buyens2014,Buyens2015a,Buyens2015,Banuls2016a,Banuls2016b,Zapp2017,Buyens2017,Banuls2018a}, and the model has been studied with a non-zero $\theta$-term~\cite{Byrnes2002,Buyens2017,Funcke2019,Angelides2022,Zache:2021ggw}, a non-zero chemical potential~\cite{Banuls2016a,Banuls2016b,Banuls2016c}, a non-zero temperature~\cite{Saito2014,Banuls2015,Saito2015,Buyens2016,Banuls2016,Banuls2018a}, and for real-time problems~\cite{Buyens2014,Buyens2016b}. The MPS approach allows us to compute the electric field and the analog of the pion condensate in a regime that is inaccessible with conventional MC methods. Moreover, MPS allow for direct access to the entanglement structure in the state, which allows us to characterize the Dashen phase transition in terms of the scaling of the correlations, thus obtaining insights into the order of the transition.

The paper is structured as follows. In Sec.~\ref{sec:model_methods}, we briefly introduce the Hamiltonian lattice formulation with staggered fermions for multiple flavors. Furthermore, we explain the numerical MPS techniques that we use to compute the ground state of the Hamiltonian and to measure observables. In Sec.~\ref{sec:results}, we present our results of the phase structure of the model. We discuss these results and provide a conclusion Sec.~\ref{sec:outlook_discussion}.

\section{Model and Methods\label{sec:model_methods}}

In the following, we briefly introduce the Hamiltonian lattice formulation of the Schwinger model, which describes quantum electrodynamics in (1+1) dimensions,  with multiple fermion flavors and staggered fermions. Subsequently, we discuss the TN methods that we use to compute the ground state of the lattice Hamiltonian.

\subsection{Lattice Schwinger model}

For our study, we adopt a Hamiltonian formulation of the Schwinger model. In the continuum, it reads
\begin{align}
  \begin{aligned}
    H = \int dx\,\Bigl( \sum_f\bigl[\bigr.&-i\overline{\psi}_f\left(\partial_1 - igA_1\right)\psi_f+ m_f\overline{\psi}_f\psi_f\bigl.\bigr]\Bigr. \\
     &\Bigl. + \frac{1}{2}E^2\Bigr),
  \end{aligned}
  \label{eq:Hamiltonian_continuum}
\end{align}
where $\psi_f$ is a two-component spinor describing a fermion of flavor $f$, $A_\mu$ is the U(1) gauge field, and we have chosen temporal gauge, $A_0=0$. The parameter $g$ is the bare coupling, and $m_f$ is the bare mass for flavor $f$. The operator $E=-\dot{A}^1$ represents the electric field, and we have used the convention $\gamma^0 = \sigma^z$ and $\gamma^1 = i\sigma^y$ for the Dirac matrices with $\sigma^{y,z}$ the usual Pauli matrices. In addition, physical states have to fulfill Gauss law 
\begin{align}
  \partial_1 E = g\sum_f\overline{\psi}_f\gamma^0\psi_f.
  \label{eq:gauss_law_continuum}
\end{align}

In order to numerically simulate the model with MPS, we work with a lattice version of the continuum Hamiltonian in Eq.~\eqref{eq:Hamiltonian_continuum}. While it has been recently shown how tackle the lattice discretization using Wilson fermions with MPS~\cite{Angelides2022}, we choose to work with staggered fermions that have been widely used in previous numerical simulations with TN~\cite{Banuls2013,Buyens2013,Buyens2015,Banuls2016a,Buyens2017,Funcke2019,Silvi2019,Felser2019,Magnifico2021}. On a lattice with spacing $a$ and $N$ sites, the lattice Hamiltonian for $F$ fermion flavors reads~\cite{Kogut1975} 
\begin{align}
  \begin{split}
  H = & -\frac{i}{2a}\sum_{n=0}^{N-2}\sum_{f=0}^{F-1}\left(\phi^\dagger_{n,f}e^{i\theta_n}\phi_{n+1,f}-\text{h.c.}\right)\\
  &+\sum_{n=0}^{N-1}\sum_{f=0}^{F-1}(-1)^nm_f\phi^\dagger_{n,f}\phi_{n,f}+ \frac{g^2a}{2}\sum_{n=0}^{N-2} L_n^2.
  \end{split}
  \label{eq:Hamiltonian}
\end{align}
Here, the operators $\phi_{n,f}$ describe a single-component fermionic field of flavor $f$ on site $n$, and $L_n$ corresponds to the (dimensionless) electric field operator acting on the link between sites $n$ and $n+1$. The operator $\theta_n$ is the canonical conjugate of $L_n$ with $[\theta_n,L_{n'}]=i\delta_{nn'}$, and $\theta_n$ is restricted to $[0,2\pi)$, as we choose to work with a compact formulation. The exponential of $\theta_n$, $e^{i\theta_n}$, acts as a lowering operator for the electric flux on the link joining the sites $n$ and $n+1$. On the lattice, the Gauss law constraint from Eq.~\eqref{eq:gauss_law_continuum} translates to
\begin{align}
  L_n - L_{n-1} = Q_n,
  \label{eq:gauss_law}
\end{align}
where $Q_n = \sum_{f=0}^{F-1}\left(\phi_{n,f}^\dagger\phi_{n,f}-\frac{1}{2}\left[1-(-1)^n\right]\right)$ is the staggered fermionic charge. 

For open boundary conditions, a recursive application of Eq.~\eqref{eq:gauss_law} allows us to reconstruct the electric field purely from the fermionic charge content, after fixing the value $l_{-1}$ of the electric field on the left boundary, $L_n = \sum_{k\leq n} Q_k + l_{-1}$. Inserting this into the Hamiltonian in Eq.~\eqref{eq:Hamiltonian} and applying a residual gauge transformation allows us to fully remove the gauge fields. After making the resulting expression dimensionless, we obtain the final dimensionless lattice Hamiltonian~\cite{Banuls2013,Banuls2016a,Banuls2016b,Funcke2019}
\begin{align}
  \begin{split}
  W = &-ix\sum_{n=0}^{N-2}\sum_{f=0}^{F-1}\left(\phi^\dagger_{n,f}\phi_{n+1,f}-\mathrm{h.c.}\right)\\
  &+\sum_{n=0}^{N-1}\sum_{f=0}^{F-1}(-1)^n \mu_f\phi^\dagger_{n,f}\phi_{n,f}
  + \sum_{n=0}^{N-2} \left( \sum_{k=0}^nQ_k\right)^2,
  \end{split}
  \label{eq:hamiltonian_dimensionless}
\end{align}
where the dimensionless constant $x=1/(ag)^2$ corresponds to the inverse lattice spacing squared in units of the coupling, and $\mu_f = 2\sqrt{x}m_f/g$ is proportional to the mass in units of the coupling.

In order to investigate the Dashen phase in the Schwinger model, we focus on the simplest nontrivial setup and restrict ourselves to two fermion flavors. In the case of QCD with two fermion flavors, the onset of the Dashen phase is characterized by the formation of a pion condensate, given by $\langle\overline{\psi}(x)\gamma^5\tau_3\psi(x)\rangle$ in the continuum theory. Here, $\psi(x)$ is a spinor having both flavor and Dirac indices, and $\tau_3$ acts on flavor space and corresponds to the Pauli matrix $\sigma^z$. The lattice analog of the pion condensate in the Schwinger model with the staggered fermion formulation reads
\begin{align}
  C=i\frac{\sqrt{x}}{N}\sum_{n=0}^{N-2}\sum_{f=0}^{1}(-1)^{n+f}\left(\phi^\dagger_{n,f}\phi^\dagger_{n+1,f}-\text{h.c.}\right)
\end{align}
in units of the coupling. Furthermore, in our study we compute the average electric field, which is given by
\begin{align}
  \bar{F} = \frac{1}{k}\sum_{n=N/2-k/2+1}^{N/2+k/2} L_n,
  \label{eq:average_field}
\end{align}
where we sum over $k$ sites in the center of the system to avoid boundary effects~\footnote{In the staggered formulation, the components of each Dirac spinor are distributed to two distinct lattice sites, thus we always choose even values for $N$. Without loss of generality we also use even values for $k$, for odd values of $k$ the summation boundaries in Eq.~\eqref{eq:average_field} have to be adjusted accordingly to obtain valid site indices.}.

\subsection{Matrix product states}

In order to numerically explore the phase structure of the Schwinger model Hamiltonian in Eq.~\eqref{eq:hamiltonian_dimensionless}, we use MPS techniques. MPS are a family of entanglement-based ansätze for the wave function of a (strongly-correlated) quantum many-body system. For open boundaries and $N$ sites, the ansatz is given by 
\begin{align}
  \ket{\chi} = \sum_{i_0,\dots,i_{N-1}}M^{i_0}_0 \dots M^{i_{N-1}}_{N-1} \ket{i_0}\otimes\dots\otimes \ket{i_{N-1}},
  \label{eq:mps}
\end{align}
where $\{\ket{i_k}\}_{k=0}^{d-1}$ is a local basis for the Hilbert space of site $k$, $M^{i_k}_k$ are complex $D\times D$ matrices for $0<k<N-1$, and $M^{i_0}_0$ ($M^{i_N}_N$) is a $D$-dimensional row (column) vector. The size of the matrices $M^{i_k}_k$, called the bond dimension of the MPS, determines the number of parameters in the ansatz. For a fixed value of the bond dimension, an MPS approximation for the ground state can be obtained variationally by iteratively optimizing the parameters until convergence~\cite{Schollwoeck2011,Orus2014a,Bridgeman2017}. After obtaining the ground state from the variational procedure, we can measure the electric field and the pion condensate.
Furthermore, MPS allow for easy access to the reduced density operator $\rho_\text{sub}$ describing a contiguous subsystem of the entire system~\cite{Schollwoeck2011,Orus2014a,Bridgeman2017}. In turn, this allows for easy access to the von Neumann entropy $S=-\tr(\rho_\text{sub}\log\rho_\text{sub})$, which is a measure for the quantum correlations between the subsystem and its environment.

For convenience, we choose to translate the fermionic degrees of freedom in Eq.~\eqref{eq:hamiltonian_dimensionless} to spins using a Jordan-Wigner transformation~\cite{Hamer1997,Banuls2013,Banuls2016a,Banuls2017,Funcke2019}. We note that MPS and more general TN are, however, able to directly deal with fermionic degrees of freedom with essentially no additional cost in the algorithm~\cite{Pineda2010,Corboz2010}.

Numerical techniques based on MPS and more general TN do not suffer from the sign problem, and allow for reliable computations in parameter regimes which are inaccessible with MC methods~(see, e.g., Refs.~\cite{Banuls2019,Felser2019,Magnifico2021,Nakayama2021,Angelides2022}). In particular, we can access the Schwinger model in the regime of negative fermion mass (or equivalently in the regmine of a topological $\theta$-term with $\theta=\pi$), which allows us to explore the Dashen phase transition in the two-flavor Schwinger model.

\section{Results\label{sec:results}}

In order to investigate the Dashen phase in the two-flavor Schwinger model, we fix the bare mass of the first flavor, $m_0/g$, to a positive value of $0.25$ and scan the bare mass of the second flavor, $m_1/g$, around $-m_0/g$. Note that this parameter regime would in general lead to a sign problem for MC methods. In our study, we use fixed dimensionless physical volumes $N/\sqrt{x}$ ranging from $10$ to $25$, with lattice spacings corresponding to $x\in[60;100]$. Moreover, we focus on the sector of vanishing total charge, $\sum_n Q_n = 0$~\footnote{In our numerical computations, vanishing total charge is imposed by adding a positive semidefinite penalty term $\lambda (\sum_n Q_n)^2$ to the Hamiltonian with sufficiently large positive constant $\lambda$. For all our calculations, we checked that the expected value of the total charge is zero to numerical precision.}. In order to estimate the error due to the finite bond dimension in our numerical simulations, we repeat the computation for every combination of $(N/\sqrt{x}, x, m_0/g, m_1/g)$ for multiple values of $D\in[20;300]$ and extrapolate to the limit $D\to \infty$ following Ref.~\cite{Funcke2019}.

\begin{figure*}[htp!]
  \centering
  \includegraphics[width=0.98\textwidth]{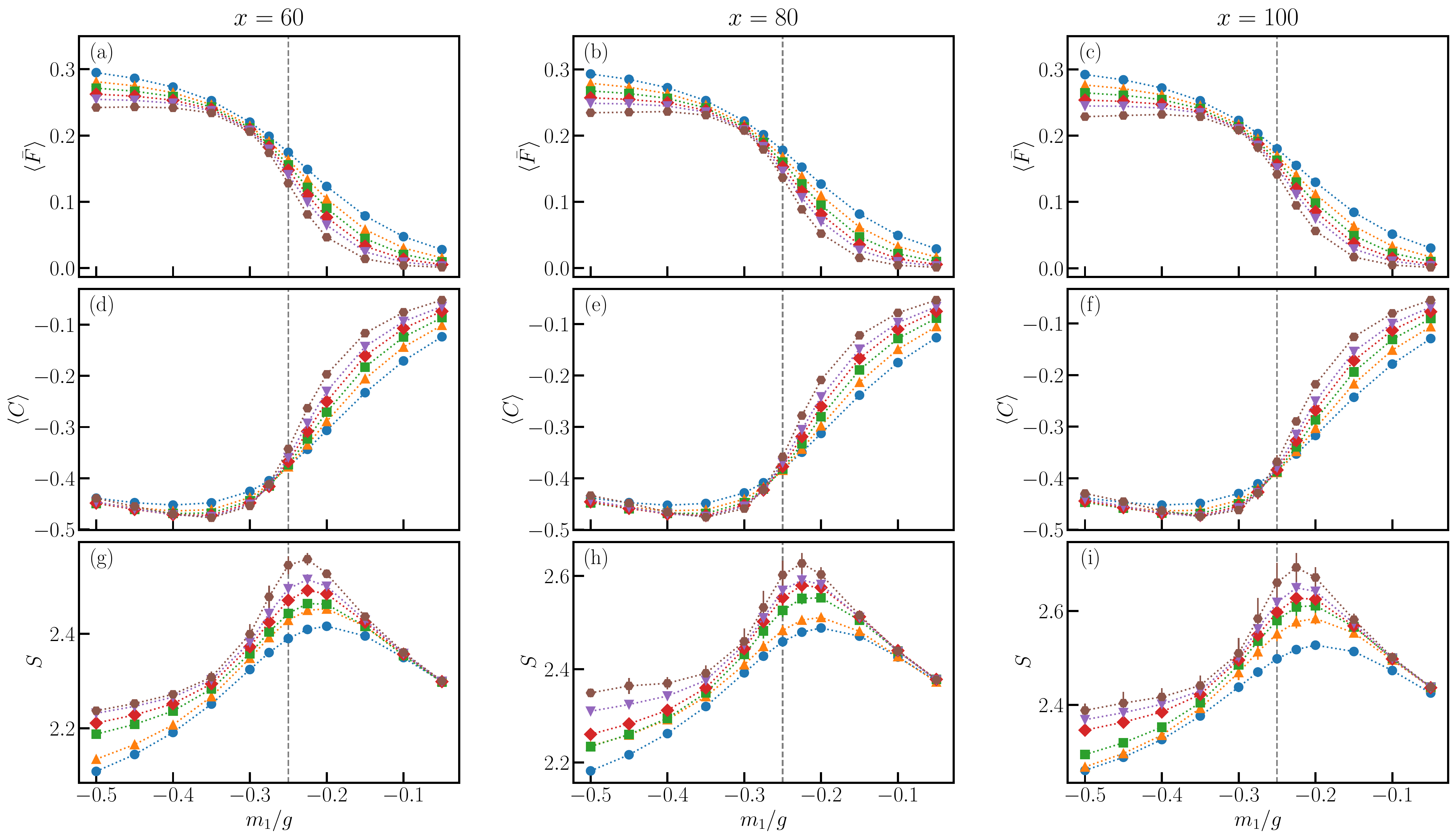}
  \caption{Average electric field (first row), pion condensate (second row) and entropy (third row) as a function of the bare mass $m_1/g$ of the second fermion flavor, for $m_0/g=0.25$, $x=60$ (first column), $x=80$ (second column), and $x=100$ (third column). Different markers indicate data for different volumes with $N/\sqrt{x}=10$ (blue dots), $12.5$ (orange triangles), $15$ (green squares), $17.5$ (red diamonds), $20$ (purple upside-down triangle), and $25$ (brown hexagons). The error bars arise from the extrapolation in the bond dimension $D$. The dashed vertical line indicates the point where $m_2/g$ reaches $-m_1/g$. To compute the average electric field, we use $k=4$ sites in the center of the system, according to Eq.~\eqref{eq:average_field}.}
  \label{fig:res_extD}
\end{figure*}

Figure~\ref{fig:res_extD} shows our results after this extrapolation for the average electric field, the pion condensate and the von Neumann entropy for various lattice spacings.
Focusing on the average electric field $\langle \bar{F}\rangle$, Figs.~\ref{fig:res_extD}(a)-(c) demonstrate that $\langle \bar{F}\rangle$ increases as we decrease the value of $m_1/g$ from above to below $-m_0/g$. While for our smallest volume, $N/\sqrt{x}=10$, this change is somewhat continuous, it becomes sharper with increasing volume. For the largest volume of $N/\sqrt{x}=25$, we observe two plateaus: for large values of $m_1/g \gg m_0/g$, the average electric field $\langle \overline{F}\rangle$ approaches zero, while for small values of $m_1/g=-0.5\ll m_0/g$, the average electric field saturates at around $0.22$. Comparing the results with different lattice spacings, we do not observe a strong dependence on $x$ throughout the entire parameter regime we study. In particular, the values for the average electric field obtained for $x=80$ and $x=100$ are essentially the same, as shown in Figs.~\ref{fig:res_extD}(b) and \ref{fig:res_extD}(c).

Due to the chiral anomaly, a negative fermion mass is equivalent to the presence of a topological $\theta$-term with $\theta=\pi$. The Schwinger model with a single fermion flavor is known to undergo a phase transition at this point, provided that the bare fermion mass is above a critical value given by $(m/g)_c\approx 0.33$ in the continuum~\cite{Byrnes2002,Buyens2017}. Previous numerical studies observed that the electric field vanishes for mass values below the phase transition point, whereas the electric field acquires a nonvanishing value after the transition occurs~\cite{Byrnes2002,Buyens2017}. Our results for the two-flavor case are compatible with this observation, and the drop in the average electric field hints towards a phase transition at $m_1/g\approx -m_0/g$.

The results for the pion condensate $\langle C \rangle$ in Figs.~\ref{fig:res_extD}(c)--(f) show a qualitatively similar picture. For large values of $m_1/g$, the value of the condensate is close to zero. When decreasing the value of $m_1/g$, we see a decrease in $\langle C \rangle$ as we approach $-m_0/g$. Upon further lowering the value of $m_1/g$, the pion condensate eventually approaches a constant value. Also for this observable, the lattice effects are small, and there is virtually not dependence on the value of $x$ throughout the parameter regime we consider. Compared to the electric field, we observe slightly larger finite-volume effects, especially in the region where $m_1/g$ is larger than $-m_0/g$. In particular, the expectation value of the pion condensate becomes closer to zero with increasing $N/\sqrt{x}$ but does not reach zero for the volumes we study.

Our observations for the pion condensate are compatible with the theoretical expectation for the Dashen phase in QCD with two fermion flavors, as presented in Sec.~\ref{sec:introduction} and Fig.~\ref{fig:PhaseDiagram}. Away from the Dashen phase, i.e., as long as  $m_1/g\gg -m_0/g$, the values for the pion condensate are close to zero. Approaching the point $m_1/g\approx -m_0/g$, at which we expect the onset of the Dashen phase, the values of the condensate abruptly decrease. When continuing to decrease the mass of the second flavor until $m_1/g\ll -m_0/g$, we eventually reach a point where $\langle C \rangle$ is roughly constant. Thus, the behavior of the condensate provides a strong indication for the occurrence of the Dashen phase transition at $m_1/g\approx -m_0/g$.

\begin{figure*}[t!]
  \centering
  \includegraphics[width=0.98\textwidth]{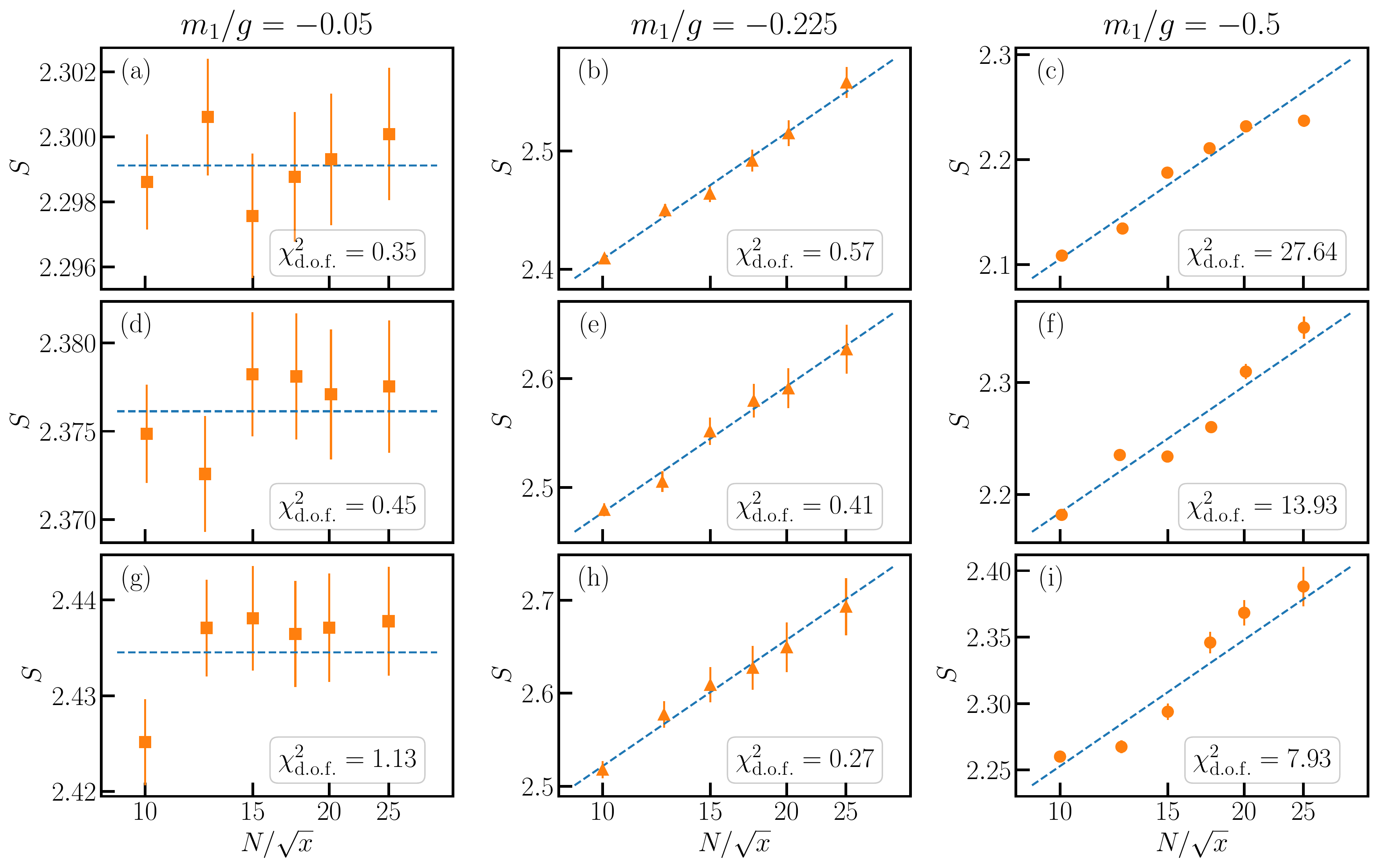}
  \caption{Entropy as a function of the volume for $x=60$ (top row), $x=80$ (middle row), and $x=100$ (bottom row). The different columns correspond to $m_1/g=-0.05$ (first column), $m_1/g=-0.225$ (middle column), and $m_1/g=-0.5$ (right column). Note that the $x$-axes are in logarithmic scale. The blue dashed lines in panels (a), (d), and (g) correspond to a constant fit, whereas in on all other panels it corresponds to a fit $S(N/\sqrt{x}) = c_1\log(N/\sqrt{x}) + c_2$ with constants $c_1$, $c_2$. The insets show the reduced $\chi^2_\mathrm{d.o.f.}$ of the fit, which demonstrate that the data are compatible with the fit function in all cases except for panels (c), (f), and (i). Error bars come from the extrapolation in $D$.}
  \label{fig:entropy_vs_volume}
\end{figure*}

Finally, we compute the von Neumann entropy $S$, as shown in Figs.~\ref{fig:res_extD}(g)--(i). Similar to the average electric field and the pion condensate, the entropy also shows a feature of the phase transition in form of a peak around $m_1/g\approx -0.225$. Comparing the results for $S$ to the other observables, we see that both finite-lattice and finite-volume effects are significantly stronger, especially around the point $m_1/g\approx -m_0/g$. The dependence of the peak value of the entropy on the volume is expected for a phase transition of second or higher order. The von Neumann entropy is a direct measure for the correlation length in the system, which in turn shows a logarithmic divergence as one approaches the phase transition point (for a system in the thermodynamic limit)~\cite{Vidal2003a,Latorre2004,Calabrese2009}. 
As we consider finite volumes, the system size bounds the correlation length from above.
Thus, for a fixed value of $x$, we expect the entropy to scale as $S\propto\log(N/\sqrt{x})$ at the transition point. In contrast, as one goes away from the critical point, the correlation length is finite, and $S$ should approach a constant value, as soon as the volume is significantly larger than the correlation length. 

With our MPS data for the entropy, we can examine this behavior. Figure~\ref{fig:entropy_vs_volume} shows the entropy at fixed values of $m_1/g$ as a function of the volume, $N/\sqrt{x}$.
Focusing on the left column in the figure, which corresponds to our largest mass, $m_1/g=-0.05$, we see that the entropy essentially reaches a constant value, independent of the volume, as expected. 
Looking at the opposite limit, $m_1/g=-0.5$, as shown in the right column in Fig.~\ref{fig:entropy_vs_volume}, our data for the entropy seems to be compatible with a logarithmic divergence at first sight. However, a closer look reveals that this is not the case. Inspecting the results for our coarsest lattice spacing, corresponding to $x=60$, in Fig.~\ref{fig:entropy_vs_volume}(c), for which our numerical data are most precise, we observe that the entropy changes its scaling behavior around volumes $12.5$ and $20$, where at the latter value it shows a trend towards saturating. Fitting our data to a logarithmic divergence, we obtain a large value for $\chi^2_\mathrm{d.o.f.}$, thus confirming that our data is not compatible with this functional form. This indicates that the correlation length is finite and, upon reaching large enough volumes to accommodate the correlation length, the entropy eventually saturates.
Going to finer lattice spacings corresponding to $x=80$ and $x=100$, we see a similar behavior, as Figs.~\ref{fig:entropy_vs_volume}(f) and \ref{fig:entropy_vs_volume}(i) reveal. Again, we observe a change in scaling behavior around the volumes $12.5$ and $20$. For these finer lattice spacings, we do not see the saturation effect as clearly as before. This can be explained by the fact that for finer lattice spacings, the correlation length in units of the lattice spacing increases; thus, one needs larger volumes to see the saturation effect. Fitting our data to a logarithmic divergence as before, the values of $\chi^2_\mathrm{d.o.f.}$ are again significantly larger than $1$, hence showing that the data is incompatible with the fit function.

Turning to $m_1/g=-0.225$, which is approximately the value for the mass at which the entropy showed a pronounced peak for all values of $x$ we study, we observe that $S$ clearly follows a logarithmic divergence in the volume, as the central column of Fig.~\ref{fig:entropy_vs_volume} demonstrates. In particular, compared to the previous cases, there is no noticeable change in the scaling behavior throughout the entire range of volumes we study, and the data is well described by a logarithmic divergence for all our lattice spacings. This is also confirmed by fitting our results to a function of the form $S(N/\sqrt{x}) = c_1 \log(N/\sqrt{x}) + c_2$, which yields values for $\chi^2_\mathrm{d.o.f.}$ below one, thus showing that the data is following the expected scaling behavior for a phase transition of second or higher order~\cite{Vidal2003a,Latorre2004,Calabrese2009}. 

Our data for the entropy corroborate the picture obtained from the electric field and the pion condensate. We observe clear indications for the occurrence of a second or higher-order phase transition, manifesting itself in a distinct peak in the entropy. While the electric field and the pion condensate only give an indication for the regime of the mass values in which the transition occurs, the peak in the entropy is a lot sharper and allows us to clearly determine the location of the transition within our resolution at $m_1/g\approx -0.225$. Interestingly, this value of the bare fermion mass is slightly larger than $-m_0/g$, which is in accordance with the theoretical expectation for the two-flavor QCD presented in Sec.~\ref{sec:introduction}, see Fig.~\ref{fig:PhaseDiagram}. This shift of the critical mass away from $-m_0/g$ does not change when varying the lattice spacing or the volume for sufficiently large volumes (see Fig.~\ref{fig:res_extD}), which indicates that the shift is a physical effect just as in QCD.

\section{Discussion and Outlook\label{sec:outlook_discussion}}

In this study, we used numerical techniques based on MPS to explore the Dashen phase in the Schwinger model with two fermion flavors. Fixing the first bare fermion mass $m_0/g$ to a positive value and scanning the value of the second bare fermion mass $m_1/g$ around $-m_0/g$, we explore a regime in which conventional MC methods suffer from the sign problem. Our numerical results indicate the existence of phase transition, which manifests itself in a sudden change in the value of the electric field as well as the formation of a pion condensate as we decrease $m_1/g$ below $-m_0/g$. The formation of the CP-violating condensate confirms that the transition is the analog of the CP-violating Dashen phase transition in QCD, and our results are compatible with the qualitative picture for QCD with two fermion flavors~\cite{Creutz:1995wf,Creutz2013,Creutz2014,Creutz2019}.

With our MPS approach, we have direct access to the bipartite entanglement in the system. Studying the the von Neumann entropy, we observe a clear peak in the entropy for a value of $m_1/g\approx -0.225$ slightly larger than $-m_0/g=-0.25$, indicating that the transition occurs slightly before the point where the absolute values of both masses are equal. This is again in agreement with the two-flavor QCD picture, for which corrections  $\propto (m_u-m_d)^2$ shift the Dashen phase transition to larger values of the down-quark mass~\cite{Gasser1983,Gasser1984,Creutz:1995wf,Creutz2013,Creutz2014,Creutz2019}.

In addition, the scaling behavior of the entanglement entropy allows for obtaining insights into the order of the transition. Studying the scaling with the volume at a fixed value of the lattice spacing, we observe a logarithmic divergence at $m_1/g=-0.225$, whereas away from the transition point the entanglement entropy does not follow this behavior. These results suggest that the observed transition is of second (or higher) order, in agreement with the theoretical prediction for QCD that two first-order transition lines with second-order end points should exist along the $m_d$ axis~\cite{Creutz:1995wf}, which has not been verified with numerical simulations so far.

In our study, we focused on negative masses, which are equivalent to having a topological $\theta$ term with $\theta=\pi$. This regime can also be addresses with TN~\cite{Byrnes2002,Buyens2017,Funcke2019,Angelides2022} and could provide an alternative route to explore the Dashen phase transition. Moreover, with recent developments for TN in (3+1) dimensions~\cite{Magnifico2021}, it might be possible to directly study QCD with two fermion flavors with TN techniques in the future. In addition, our results can also serve as a benchmark for recent efforts to simulate lattice gauge theories on quantum hardware~\cite{Clemente2022,Nguyen2022,Funcke2022a,Thompson2022}.

\acknowledgments
S.K.\ acknowledges financial support from the Cyprus Research and Innovation Foundation under projects ``Future-proofing Scientific Applications for the Supercomputers of Tomorrow (FAST)'', Contract No.\ COMPLEMENTARY/0916/0048 and ``Quantum Computing for Lattice Gauge Theories'', Contract No.\ EXCELLENCE/0421/0019.
L.F.\ is partially supported by the U.S.\ Department of Energy, Office of Science, National Quantum Information Science Research Centers, Co-design Center for Quantum Advantage (C$^2$QA) under Contract No.\ DE-SC0012704, by the DOE QuantiSED Consortium under Subcontract No.\ 675352, by the National Science Foundation under Cooperative Agreement No.\ PHY-2019786 (The NSF AI Institute for Artificial Intelligence and Fundamental Interactions, \url{http://iaifi.org/}), and by the U.S.\ Department of Energy, Office of Science, Office of Nuclear Physics under Grants No.\ DE-SC0011090 and No.\ DE-SC0021006.
\bibliography{Papers}
\end{document}